\documentclass{ws-procs975x65}
\usepackage{graphicx}
\usepackage{amsmath}
\usepackage{amssymb}
\usepackage{enumerate}
\usepackage{epsfig}
\usepackage{verbatim}
\usepackage{array}

\makeatletter

\def\vereq#1#2{\lower3pt\vbox{\baselineskip0.5pt\lineskip0.5pt
\ialign{$\m@th#1\hfill##\hfil$\crcr#2\crcr\sim\crcr}}}
\def\gtrsim{\mathrel{\mathpalette\vereq>}}
\makeatother

\newcommand{\regularwidthofpicture}{0.7\textwidth}

\newcommand{\astar}{\ensuremath{a_*}}

\newcommand{\omtil}{\ensuremath{\widetilde{\omega}}}
\newcommand{\Qtil}{\ensuremath{\widetilde{Q}}}

\newcommand{\bmax}{\ensuremath{b_\mathrm{max}}}

\newcommand{\TeV}{\ensuremath{\,\mathrm{TeV}}}

\newcommand{\fb}{\ensuremath{\,\mathrm{fb}}}
\newcommand{\pb}{\ensuremath{\,\mathrm{pb}}}

\newcommand{\mycite}[1]{[\refcite{#1}]}

\newcommand{\Sch}{Schwarz\-schild }

\begin{document}

\title{Rotating Black Holes/Rings at Future Colliders}

\author{Daisuke IDA}
\address{
  Department of Physics, Tokyo Institute of Technology, Tokyo 152-8551, Japan \\
  e-mail: d.ida@th.phys.titech.ac.jp}
\author{Kin-ya ODA\footnote{
  \uppercase{I}nvited talk in
  the 10th \uppercase{M}arcel \uppercase{G}rossmann \uppercase{M}eeting
  in \uppercase{R}io de \uppercase{J}aneiro,
  20--26 \uppercase{J}uly 2003}}
\address{
  Physik Department T30e, TU M\"unchen, D-85748 Garching, Germany \\
  e-mail: odakin@ph.tum.de}
\author{Seong Chan PARK}
\address{
  Korea Institute for Advanced Study (KIAS), Seoul 130-012, Korea \\
  e-mail: spark@kias.re.kr}

\maketitle

\abstracts{
The hierarchy between the electroweak and Planck scales can be reduced
when the extra dimensions
are compactified with large volume or with warped
geometry,
resulting in the fundamental scale
of the order of TeV.
In such a scenario, one can experimentally study
the physics \emph{above} the Planck scale.
We discuss black hole/ring production at future colliders.
}

\section{Introduction}
Black hole production is one of the most important prediction
in the large~\mycite{Arkani-Hamed:1998rs} and warped (RS1)~\mycite{Randall:1999ee} extra dimension scenarios
in which the fundamental gravitational scale becomes of the order of TeV.\footnote{
  In RS1 scenario the fundamental scale changes along
  the extra dimension, being of the order of TeV at our visible brane.}
The classical black hole production cross section
in higher dimensions
(for initial two point-particles stuck on the 3-brane)
is roughly~\mycite{Giddings:2001bu,Dimopoulos:2001hw}
\begin{align}
  \sigma &\sim \pi r_S^2
          \sim {1\over M_P^2}\left(s\over M_P^2\right)^{1 \over n+1},
		  \label{schwarzschild_eq}
  \end{align}
where $n$~is the number of extra dimensions,
$M_P$~is
the higher dimensional Planck scale,
and $r_S$~is the Schwarz\-schild radius
of the higher dimensional black hole whose
mass is equal to the center-of-mass (c.m.) energy $\sqrt{s}$
(see also Refs.~\mycite{Yoshino:2002tx,Banks:1999gd}).
Note that this cross section \emph{increases}
with the c.m.\ energy\footnote{
  We note that this process is truly non-perturbative and
  that there is no contradiction with the argument of
  the perturbative unitarity in local quantum field theory.}
and therefore this process will eventually
dominate over \emph{any} short distance interactions
as one increases the energy above~$M_P$.\footnote{
  Recently, the opposite possibility is proposed that
  the gravity becomes \emph{weak} rather than strong
  above its cut-off scale set at~$\sim 10^{-3}\mathrm{eV}$,
  though it is yet unknown how to realize it
  in some quantum gravity model (or in string theory)~\mycite{Sundrum:2003tb}.}
In ordinary four dimensional ($n=0$) gravity,
this cross section gives $\sigma \sim 10^{-50}\fb$
even for $\sqrt{s}=100\TeV$ and does not seem accessible
within our current or near future technology.
In contrast, in the presence of large/warped extra dimension(s)
the Planck scale~$M_P$ is of order TeV and
the cross section \eqref{schwarzschild_eq} gives
$\sigma\gtrsim\TeV^{-2} \simeq 400\pb$
at the typical energy scale of the CERN Large Hadron Collider (LHC)
leading to millions of black holes per year~\mycite{Dimopoulos:2001hw}.

The black hole production process is not only interesting to consider
but also fundamental in the following sense.
Well above the TeV scale, all the sub-processes with shorter
length scale than $O(\mathrm{TeV}^{-1})$ are hidden by the event
horizon of the black hole
and hence
the \emph{only} things we can
observe are black holes and their
decay products~\mycite{Giddings:2001bu}.
This situation is the same in string theory.
For a fixed string coupling~$g_s$
only the cross section below the energy scale $g_s^{-2}M_s$
can be calculated within its perturbative framework,
where $M_s=1/\sqrt{\alpha'}$ is the string scale ($\simeq M_P$
for $g_s\simeq 1$).
Above this scale, the string perturbation theory breaks down
and the string picture is expected to be altered by
the black hole picture
in which the semi-classical treatment becomes
better and better as one increases the energy.
(Around this scale, a black hole is 
related to massive modes of a single string via
the correspondence principle~\mycite{Dimopoulos:2001qe}.\footnote{
  Originally the argument of the correspondence principle
  is for fixed energy (mass),
  varying the string coupling~\mycite{Horowitz:1997nw},
  but the same argument holds
  for fixed coupling varying the energy (mass).})
This type of infrared-ultraviolet (IR-UV) duality always 
appears when one tries to obtain a quantum description of
gravitational interactions:
One can describe the IR region perturbatively, which can be
mapped into the UV region by a duality.
The region of true interest is intermediate one where
both pictures break down and a non-perturbative formulation of quantum
gravity (or string theory) becomes relevant.
Given the status of the theoretical development,
an experimental signature of quantum gravity in this intermediate region
would be observed as discrepancy
from the semi-classical behavior in the black hole picture,
which is universal in the high energy limit.
Therefore in order to investigate quantum gravity effects,
it is essential to predict the semi-classical black hole behavior
as precisely as possible.
This is the main motivation of our work~\mycite{Ida:2002ez}.

\section{Production}
\subsection{Black holes}
\begin{figure}[tbhp]
  \begin{center}
    \vspace{-0.5cm}
	\leavevmode \epsfig{file=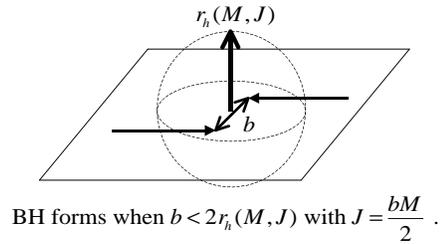,width=\regularwidthofpicture}
    \vspace{-2cm}
    \caption{
      Schematic picture for the condition of the black hole formation.
      \label{fig:BHpicture} }
    \end{center}
  \end{figure}

Following the above argument,
we assume that the classical black hole production cross section
is a good approximation for 
the collision of two partons with $\sqrt{s}$ sufficiently larger than $M_P$.
Let us consider two massless particles colliding with impact parameter $b$
and c.m.\ energy $\sqrt{s}=M_i$
so that each particle has momentum $M_i/2$ in the c.m.\ frame.
Neglecting the spins of colliding particles,
the initial angular momentum before collision is $J_i=bM_i/2$.
(See Fig.~\ref{fig:BHpicture} for a schematic picture.)
Suppose that a black hole forms
whenever the initial two particles (characterized by $M_i$ and $J_i$)
can be wrapped inside the event horizon of
the black hole with the mass $M=M_i$ and angular momentum $J=J_i$:
\begin{align}
  b &< 2r_h(M,J)=2r_h(M_i,bM_i/2),
  \label{eq:our_condition}
  \end{align}
where $r_h(M,J)$ is
the horizon radius of the higher dimensional Kerr black hole.

Since $r_h(M,J)$ is a monotonically decreasing function of $J$ for fixed $M$,
there is a maximum impact parameter $\bmax$ which saturates the
inequality~(\ref{eq:our_condition}):
\begin{align}
  \bmax(M) &= 2\left[1+\left(\frac{n+2}{2}\right)^2\right]^{-{1\over n+1}}r_S(M),
  \label{eq:bmax_formula}
  \end{align}
giving the black hole production cross section $\sigma=\pi\bmax^2$.
The formula (\ref{eq:bmax_formula})
fits the numerical result of $\bmax$
with full consideration of general relativity
by Yoshino and Nambu~\mycite{Yoshino:2002tx}
within an accuracy of
less than 1.5\% for $n\geq 2$ and 6.5\% for $n=1$.

We note that this result is obtained in the 
approximation where we neglect all the effects of emissions
during the formulation of the black hole (balding phase),
\textit{i.e.} we assume that 
the initial c.m.\ energy $M_i$ and angular momentum $J_i$ become directly
the resultant black hole mass $M=M_i$ and angular momentum $J=J_i$.
The coincidence of our result with the numerical study
suggests that this approximation is actually viable
for higher dimensional black hole formation
at least unless $b$ is very close to $\bmax$.

Once we neglect the balding phase, 
the initial impact parameter $b$ directly leads to
the resultant angular momentum of the black hole $J=bM/2$.
Since the impact parameter $[b, b+db]$ contributes to the
cross section $2\pi bdb$,
we obtain the (differential) production cross section
of the black hole with mass $M$ and angular momentum in $[J,J+dJ]$
\begin{align}
d\sigma(M,J)
  &=\begin{cases}
      8\pi JdJ/M^2 & (J<J_\mathrm{max})\\
      0            & (J>J_\mathrm{max})
      \end{cases}, \label{eq:dsigma_dJ}
  \end{align}
where
\begin{align}
J_\mathrm{max}
   &=\frac{\bmax M}{2}=j_n\,\left(\frac{M}{M_P}\right)^{n+2\over n+1},
     \label{Jmax_eq}
\end{align}
with the values of $j_n$ summarized in Table~\ref{table:jandk}.\footnote{
  Our estimation neglecting the balding phase
  gives more or less the maximum possible values of $M$
  and $J_\mathrm{max}$.
  One can instead give the minimum possible value of $M$,
  the most conservative bound in the opposite extreme,
  utilizing the irreducible mass $M_\mathrm{A.H.}$ of
  Ref.~\mycite{Yoshino:2002tx}
  which provides a lower bound on the final mass
  of the black hole after the balding phase.
  See \textit{e.g.} Ref.~\mycite{Anchordoqui:2003jr}
  for such an analysis. \label{footnote_ref} }
\begin{table}
    \tbl{
      Maximum angular momentum \label{table:jandk}}
    {\begin{tabular}{@{}ccccccccc@{}}
      \hline
      $n$  &0     &1    &2    &3    &4    &5    &6    &7     \\
      \hline
      $j_n$    &0.0398&0.256&0.531&0.815&1.09 &1.37 &1.63  &1.88  \\
      $k_n$    &0.0159&0.125&0.228&0.251&0.214&0.155&0.101 &0.0603\\
      $k_n/j_n$&0.399 &0.489&0.429&0.308&0.195&0.114&0.0619&0.0320\\
      \hline
      \end{tabular} }
  \begin{tabnote}
    $j_n$ and $k_n$ are $J_\mathrm{max}$ and $J_\mathrm{min}$ in units of
	$(M/M_P)^{n+2\over n+1}$ required
	for black hole and black ring formations,
	respectively.
    \end{tabnote}
  \end{table}

The differential cross section~(\ref{eq:dsigma_dJ})
linearly increases with the angular momentum and 
the black hole tends to be produced with large angular momentum.
For typical mass of the black hole produced at LHC $M/M_P\sim 5$,
the value of $J_\mathrm{max}$
is $J_\mathrm{max}\sim 3, 5, \ldots, 10$ and  12 for
$n=1, 2, \ldots, 6$ and 7, respectively;
for large $n$, the angular momentum is not only sizable
but also larger than the mass in Planck units
and hence is even safer to be treated semi-classically 
than the mass is.

Integrating the expression~(\ref{eq:dsigma_dJ}), we obtain
\begin{align}
  \sigma(M)
    &=\pi\bmax^2
     =4\left[1+\left(\frac{n+2}{2}\right)^2\right]^{ -{2\over n+1} }\,\pi r_S(M)^2
	 =F\,\pi r_S(M)^2,
  \end{align}
where the form factor $F$ is summarized in Table~\ref{form_factor_table}.
\begin{table}
    \tbl{
      Comparison of analytical and numerical results
	  for form factor \label{form_factor_table}}
    {\begin{tabular}{@{}ccccccccc@{}}
      \hline
      $n$  & 0 & 1 & 2 & 3 & 4 & 5 & 6 & 7 \\
      \hline
      $F_\mathrm{Numerical}$~\mycite{Yoshino:2002tx}
        & 0.647 & 1.084 & 1.341 & 1.515 & 1.642 & 1.741 & 1.819 & 1.883\\
      $F_\mathrm{Analytic}$
        & 1.000 & 1.231 & 1.368 & 1.486 & 1.592 & 1.690 & 1.780 & 1.863\\
	  \hline
      \end{tabular} }
  \begin{tabnote}
    $F$ gives the form factor $F=\sigma/\pi r_S^2$.
    \end{tabnote}
  \end{table}
This result implies that,
apart from the four-dimensional case,
we would underestimate the production cross section of black holes
if we do not take the angular momentum into account.

\subsection{Black rings}
Black holes can have various non-trivial topologies in higher dimensions.
In particular in five dimensions ($n=1$)
one can construct an explicit solution for a stationary rotating black ring
which is homeomorphic to $S^1\times S^2$~\mycite{Emparan:2001wn}.
Here we consider the possibility of
higher dimensional $S^1\times S^{n+1}$ black ring formation.
Since we do not know how to extend
the five dimensional solution to $(4+n)$~dimensions,
we work in the Newtonian approximation for ring dynamics
assuming
that nonlinear effects will not change the qualitative features.
Let us consider a rotating massive circle
with radius~$\ell$, mass~$M$ and angular momentum~$J$
in $(4+n)$-spacetime dimensions.
For given~$J$,
the gravitational attraction~$F_g$ and centrifugal force~$F_c$ are
\begin{align}
  F_g &\sim {GM^2\over  \ell^{2+n}}, &
  F_c &\sim {J^2 \over M\ell^3    }, \label{forces_eq}
  \end{align}
where $G$ is the $(4+n)$-dimensional Newton constant.
Therefore we expect that the stationary solution is allowed only
for $n=1$ and that the ring either shrinks or explodes
monotonically for $n\geq 2$.
Let $r$ be the \Sch radius of the point mass
in the $(3+n)$-dimensional effective theory
which is obtained by integrating along the $S^1$~direction:
$r\sim (GM/\ell)^{1/n}$.
Two conditions must hold for a black ring to form
in flat space picture:
\begin{enumerate}
  \item
    $\ell>r$ must hold so that the hole of
	doughnut is not filled up;
  \item
    $F_c>F_g$
	must hold
	so that the ring does not start to shrink but to explode.
  \end{enumerate}
These conditions result in the following \emph{minimum} value
for the angular momentum:
\begin{align}
  J &\gtrsim J_\mathrm{min}=k_n\left(M\over M_P\right)^{n+2\over n+1},
  \end{align}
where the values of $k_n$ are
summarized in Table~\ref{table:jandk}.\footnote{
  Recall that the argument presented here is a qualitative
  order estimation.
  Numerical coefficients~$k_n$ are estimated by assuming 
  Newtonian forces for two point particles with each mass~$M/2$. 
  See Ref.~\mycite{Ida:2002ez} for details.
  (The value $k_0$ is meaningless but presented just for comparison.)}
This result shows that when $n$ is large,
$J_\mathrm{min}$ for exploding black rings
is one or two order(s) of magnitude 
smaller than $J_\mathrm{max}$
to form a virtual event horizon shown in Fig.~\ref{fig:BHpicture}
(\textit{i.e.} to have strong non-perturbative gravitational interactions).
Therefore we expect that the exploding black rings are possibly
produced if there are many extra dimensions,
though they will suffer from the black string instability when they become
sufficiently large thin rings and their fate is unpredictable at this stage.

\section{Evaporation}
Black holes radiate mainly on the brane~\mycite{Emparan:2000rs}.\footnote{
  One may argue that a bulk graviton emission is not negligible
  based on the following two points:
  (i)~The graviton has larger number of degrees of freedom in higher dimension;
  (ii)~The super\-radiant graviton emission for highly rotating black hole
      can be greatly enhanced by the greybody factor.
  The former point becomes milder after fixing various moduli fields.
  (Typically all the four dimensional scalars coming from
  the Kaluza-Klein decomposition of the graviton
  must be made massive for the theory to be viable.)
  The latter cannot be answered at this stage since no one has presented
  the bulk graviton field equation for the higher dimensional Kerr black hole.
  One could assume that this type of super\-radiant emission
  is already taken into account when one follows the conservative treatment
  mentioned in Footnote~\ref{footnote_ref}.
  }
So we study brane field emission from a higher dimensional black hole.

\subsection{Brane field equations}
We make the following ansatz for the Newman-Penrose null tetrads:
\begin{align}
  n
    &= dt-a\sin^2\vartheta d\varphi
       -{\Sigma\over\Delta}dr, \nonumber\\
  n'
    &= {\Delta\over 2\Sigma}\left(
	    dt-a\sin^2\vartheta d\varphi\right)
       +\frac{1}{2}dr, \nonumber\\
  m
    &= {i\sin\vartheta\over 2^{1/2}(r+ia\cos\vartheta)}\left[
         a dt-(r^2+a^2)d\varphi\right]
       -{r-ia\cos\vartheta\over 2^{1/2}}d\vartheta, \nonumber\\
  m'
    &= \bar m, \label{eq:null_tetrad}
  \end{align}
where
\begin{align}
  \Sigma &= r^2+a^2\cos^2\vartheta, &
  \Delta &= r^2+a^2-\mu r^{1-n}.
  \end{align}
It is straightforward to check that Eq.~\eqref{eq:null_tetrad}
results in the correct form of the induced four dimensional metric
(of the totally geodesic probe brane) in the higher dimensional Kerr field
\begin{align}
  g &= \left(1-\frac{\mu r^{-n+1}}{\Sigma}\right)dt^2
       -\sin^2\vartheta\left(
         r^2+a^2+a^2\sin^2\vartheta\frac{\mu r^{-n+1}}{\Sigma}
	     \right)d\varphi^2
       \nonumber\\
    &\quad +2a\sin^2\vartheta\frac{\mu
       r^{-n+1}}{\Sigma}dtd\varphi
       -\frac{\Sigma}{\Delta}dr^2-\Sigma d\vartheta^2.
	   \label{eq:brane_metric}
  \end{align}
The parameters $\mu$ and $a$ are related to the mass $M$ and angular momentum $J$
of the higher dimensional black hole by
\begin{align}
  M &= \frac{(n+2)A_{n+2}}{16\pi G}\mu, &
  J &= \frac{2}{n+2}Ma,  \label{eq:MandJ}
  \end{align}
where $A_{n+2}=2\,\pi^{n+3\over 2}/\Gamma(\frac{n+3}{2})$ is the
area of the unit sphere $S^{n+2}$.

Utilizing the null tetrads~\eqref{eq:null_tetrad}
one can show that the brane field
equations for a massless field with spin $s=0, \frac{1}{2}$ and 1 are separable
\begin{align}
  &{1\over\sin\vartheta}{d\over d\vartheta}\left(\sin\vartheta
    {d S\over d\vartheta}\right)
  +\left[
    (s-a\omega\cos\vartheta)^2
    -
    (s\cot\vartheta+m\csc\vartheta)^2
    -s(s-1)+A
    \right]S
  =0,  \label{brane_field_equation_S} \\
  &\Delta^{-s}{d\over dr}\left(\Delta^{s+1}{dR\over dr}\right)  \nonumber\\
  &\qquad+\left[
    {K^2\over\Delta}
    +s\left(
      4i\omega r
      -i{\Delta_{,r}K\over\Delta}+\Delta_{,rr}-2\right)
    -A+2ma\omega-a^2\omega^2\right]R
  =0, \label{brane_field_equation_R}
  \end{align}
where $K=(r^2+a^2)\omega-ma$, $A$ is the angular eigenvalue,
and the following decomposition is employed
\begin{align}
\Phi &= \int d\omega\,e^{-i\omega t}\sum_m e^{im\varphi}
          \sum_l R_{\omega lm}(r)S_{\omega lm}(\vartheta).
		  \label{decomposition_eq}
  \end{align}
(Here and hereafter, $m$ stands for a number in the
decomposition~\eqref{decomposition_eq} rather than
the 1-form in Eq.~\eqref{eq:null_tetrad}.)
This is one of our main results.
The angular part~\eqref{brane_field_equation_S} is not modified from four dimensions
and can be solved in terms of the spin-weighted spheroidal harmonics
${}_sS_{lm}(a\omega;\vartheta,\varphi)$
with the angular eigenvalue
\begin{align}
  A&=l(l+1)-s(s+1)-{2ms^2\over l(l+1)}a\omega+O\left((a\omega)^2\right).
  \end{align}

\subsection{Greybody factors}
The greybody factors $\Gamma$
determine the Hawking radiation for each brane mode:
\begin{align}
\frac{dN_{s,l,m}}{dt\,d\omega\,d\varphi\,d\cos\vartheta}
  &=
   \frac{1}{2\pi}
   \frac{{}_s\Gamma_{l,m}(r_h,a;\omega)}{e^{2\pi\Qtil}-(-1)^{2s}}
   \left|{}_sS_{lm}(a\omega;\vartheta,\varphi)\right|^2,
   \label{eq:power_spectrum}
  \end{align}
where $\Qtil=(\omega-m\Omega)/2\pi T$,
and hence determine the spectrum of the decay products
of the black hole completely
(up to a few quanta emitted in the Planck phase where the semi-classical
treatment of the black hole radiation breaks down).
In most literature the greybody factors are assumed to
take the form of the geometrical optics (g.o.) limit:
\begin{align}
  \Gamma_{g.o.}
    &=  \left(\frac{n+3}{2}\right)^{2\over n+1}
        \frac{n+3}{n+1}(r_h\omega)^2. \label{eq:golimit}
  \end{align}

Once we obtain the brane field equation \eqref{brane_field_equation_R},
we can calculate $\Gamma$ for each mode as the absorption probability
at infinity with purely ingoing boundary condition put at the horizon.
We define the following dimensionless quantities
\begin{align}
  \xi    &= {r-r_h\over r_h}, &
  \omtil &= r_h\omega.
  \end{align}
Then $\Qtil$ can be written as $\Qtil=(1+\astar^2)\omtil-m\astar$
where $\astar=a/r_h$.
In five dimensions ($n=1$), we solve
the radial equation~\eqref{brane_field_equation_R}
both in the near-horizon and far-field limits
$\xi \ll 1/\omtil$ and $\xi\gg 1+|\Qtil|$, respectively,
\begin{align}
  R_\mathrm{NH}
    &= \left(\frac{\xi}{2}\right)^{-s-\frac{i\Qtil}{2}}
       \left(1+\frac{\xi}{2}\right)^{-s+\frac{i\Qtil}{2}}
       {}_2F_1(-l-s,l-s+1,1-s-i\Qtil;-\frac{\xi}{2}),
	   \nonumber\\
  R_\mathrm{FF}
    &= B_1e^{-i\omtil\xi}
       \left(\frac{\xi}{2}\right)^{l-s}
       {}_1F_1(l-s+1,2l+2;2i\omtil\xi)\nonumber\\
    &\mbox{}
	  +B_2e^{-i\omtil\xi}\left(\frac{\xi}{2}\right)^{-l-s-1}
	   {}_1F_1(-l-s,-2l;2i\omtil\xi),
  \end{align}
where ${}_2F_1$ and ${}_1F_1$ are hypergeometric functions.
Then we match both solutions
at the consistent overlapping region 
$1+|\Qtil|\ll \xi \ll 1/\omtil$
in the low frequency limit $\omega\ll 1/r_h$.
By this matching we obtain the constants $B_1$ and $B_2$
in terms of $s,l$ and $\Qtil$,
which lead to the following analytic formula for the greybody factor:
\begin{align}
  \Gamma
    &= 1-\left|\frac{1-C}{1+C}\right|^2, \label{eq:Page_trick}
  \end{align}
where
\begin{align}
  C &= \frac{(4i\omtil)^{2l+1}}{4}\left(\frac{(l+s)!(l-s)!}{(2l)!(2l+1)!}\right)^2
       \left(-i\Qtil-l\right)_{2l+1}, 
  \end{align}
with $(\alpha)_n=\prod_{n'=1}^n(\alpha+n'-1)$ being Pochhammer's symbol.
For concreteness, we write down
the explicit expansion of Eq.~(\ref{eq:Page_trick}) up to $O(\omtil^6)$ terms
\begin{align}
  {}_0\Gamma_{0,0}&=4\omtil^2-8\omtil^4+O(\omtil^6), \nonumber\\
  {}_0\Gamma_{1,m}&=\frac{4\Qtil\omtil^3}{9}\left(1+\Qtil^2\right)
                    +O(\omtil^6),\nonumber\\
  {}_0\Gamma_{2,m}&=\frac{16\Qtil\omtil^5}{2025}\left(
                    1+\frac{5\Qtil^2}{4}+\frac{\Qtil^4}{4}\right)
                    +O(\omtil^{10}),\nonumber
  \end{align}\nopagebreak[4]
\begin{align}
  {}_{\frac{1}{2}}\Gamma_{\frac{1}{2},m}
    &=\omtil^2\left(1+4\Qtil^2\right)
      -\frac{\omtil^4}{2}\left(1+4\Qtil^2\right)^2
      +O(\omtil^6), \nonumber\\
  {}_{\frac{1}{2}}\Gamma_{\frac{3}{2},m}
    &=\frac{\omtil^4}{36}\left(
      1+\frac{40\Qtil^2}{9}+\frac{16\Qtil^4}{9}\right)+O(\omtil^8),
      \nonumber
  \end{align}\nopagebreak[4]
\begin{align}
  {}_1\Gamma_{1,m}
    &=\frac{16\Qtil\omtil^3}{9}\left(1+\Qtil^2\right)+O(\omtil^6), \nonumber\\
  {}_1\Gamma_{2,m}
    &=\frac{4\Qtil\omtil^5}{225}\left(
               1+\frac{5\Qtil^2}{4}+\frac{\Qtil^4}{4}\right)
               +O(\omtil^{10}). \label{eq:Gamma_explicit}
  \end{align}
We have shown
the greybody factors
for the higher dimensional Kerr black hole.
That of the \Sch black hole is included 
in the limit $\astar\rightarrow 0$
($\Qtil\rightarrow\omtil$).
Note that the low frequency behavior of vector ($s=1$)
is different from g.o.\ limit~\eqref{eq:golimit}
in its powers of $\omtil$
even in the limit $\astar\rightarrow 0$.
We can see
that the numerical coefficient quickly becomes smaller
as $l$ becomes larger.

\subsection{Radiation from Randall-Sundrum black hole}
\begin{figure}[tbhp]
  \begin{center}
    \epsfig{file=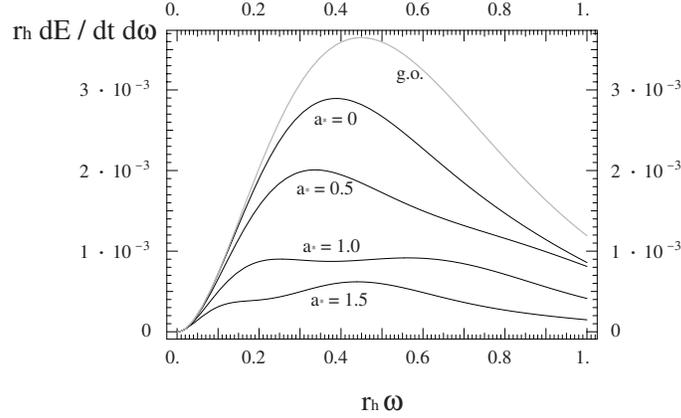,width=\regularwidthofpicture}
    \caption{
      Scalar ($s=0$) power spectrum $r_hdE/dt\,d\omega$ vs $r_h\omega$.
      The gray line is the geometrical optics limit.
      \label{fig:lins0} } 
    \end{center}
  \end{figure}

\begin{figure}[tbhp]
  \begin{center}
    \epsfig{file=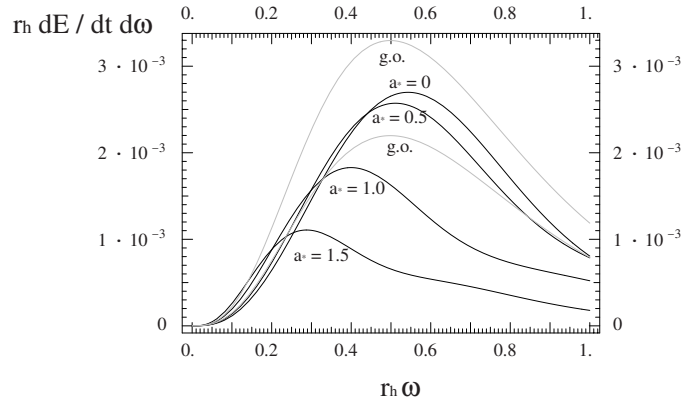,width=\regularwidthofpicture}
    \caption{
      Spinor ($s=\frac{1}{2}$) power spectrum $r_hdE/dt\,d\omega$ vs $r_h\omega$.
      Upper gray line is the geometrical optics limit
	  (lower one is multiplied by 2/3; see text). \label{fig:lins12} }
    \end{center}
  \end{figure}

\begin{figure}[tbhp]
  \begin{center}
    \epsfig{file=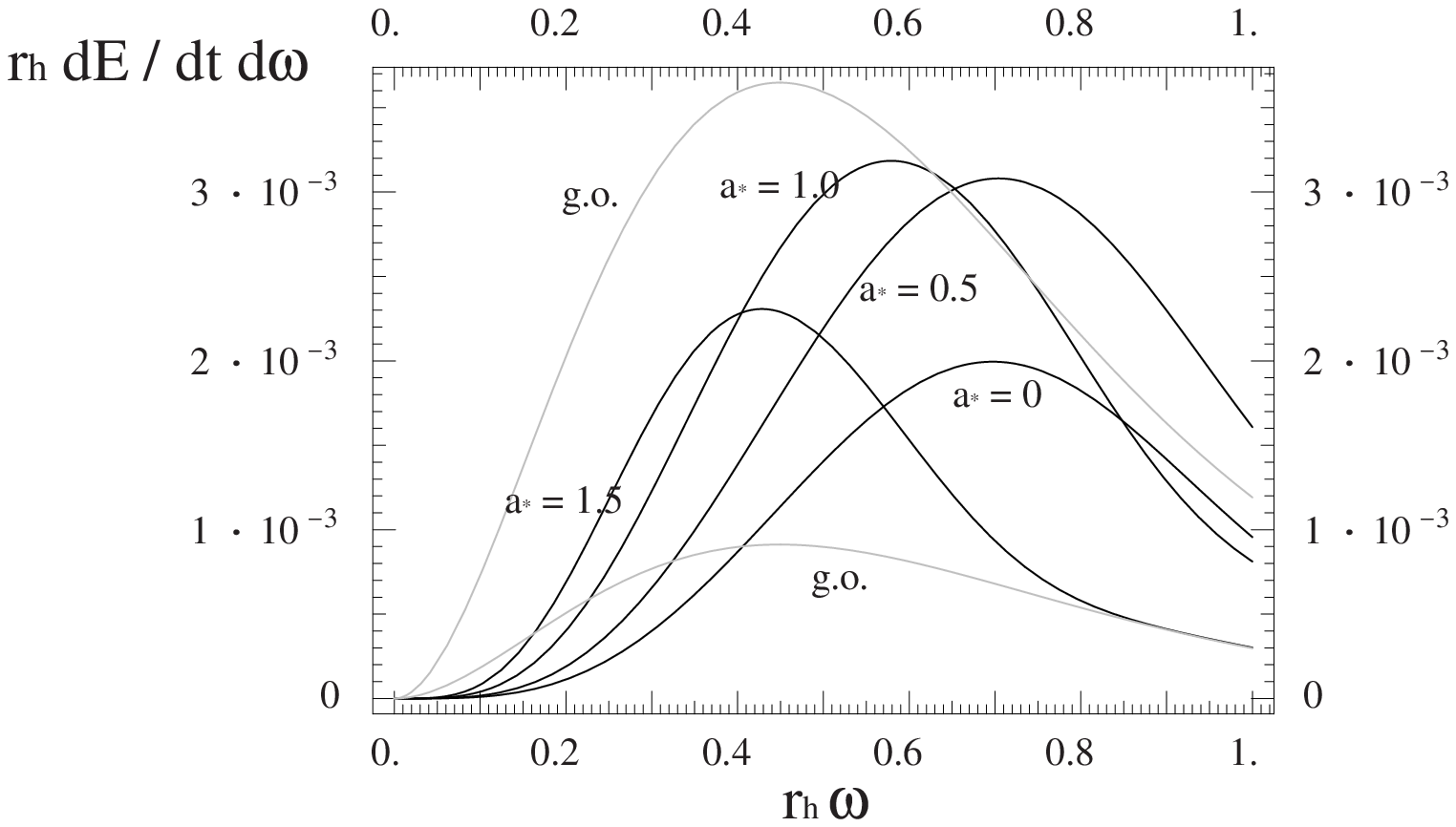,width=\regularwidthofpicture}
    \caption{
      Vector ($s=1$) power spectrum $r_hdE/dt\,d\omega$ vs $r_h\omega$.
	  Upper gray line is the geometrical optics limit
	  (lower one is multiplied by 1/4; see text). \label{fig:lins1} }
\end{center}
\end{figure}

In Figs.~\ref{fig:lins0}--\ref{fig:lins1}, we show
the power spectra for spin $s=0$, $\frac{1}{2}$ and 1
fields.
The black lines are our results from $a_*=0$ to 1.5.
(The maximum $\astar$ allowed for the black hole production
by Eq.~\eqref{Jmax_eq} is $(\astar)_\mathrm{max}=\frac{n+2}{2}=1.5$.)
Note that our approximation is valid for $\omtil<\min(1,\astar^{-1})$.
The gray line is the power spectrum
in the g.o.\ limit \eqref{eq:golimit}.
(For spinor and vector, the lower gray line is multiplied by
the phenomenological weighting factor 2/3 and 1/4, respectively,
which are introduced to mimic the four dimensional result in some papers.)

\begin{figure}[tbhp]
  \begin{center}
    \epsfig{file=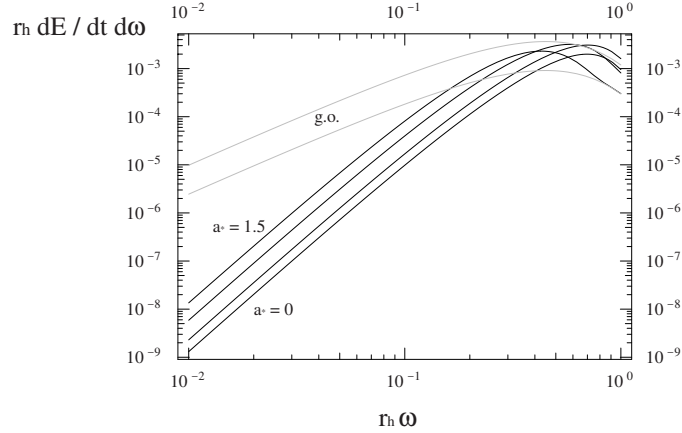,width=\regularwidthofpicture}
    \caption{
      Vector ($s=1$) power spectrum $r_hdE/dt\,d\omega$ vs $r_h\omega$
      in a log-log plot.
      Upper gray line is the geometrical optics limit
	  (lower one is multiplied by 1/4; see text).	\label{fig:logs1} } 
    \end{center}
  \end{figure}

\begin{figure}[tbhp]
  \begin{center}
    \epsfig{file=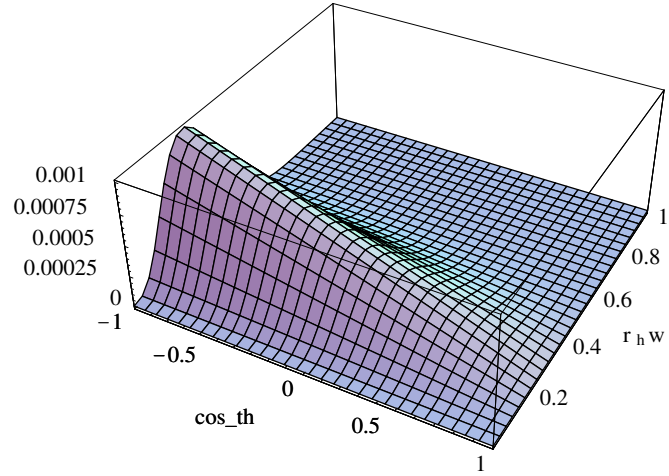,width=\regularwidthofpicture}
    \caption{
      Spinor ($s=\frac{1}{2}$)
      power spectrum $r_hdE/dt\,d\omega\,d\cos\vartheta$ vs $r_h\omega$
      and $\cos\vartheta$ for $\astar=1.5$ \label{fig:ang12} } 
    \end{center}
  \end{figure}

\begin{figure}[tbhp]
  \begin{center}
    \epsfig{file=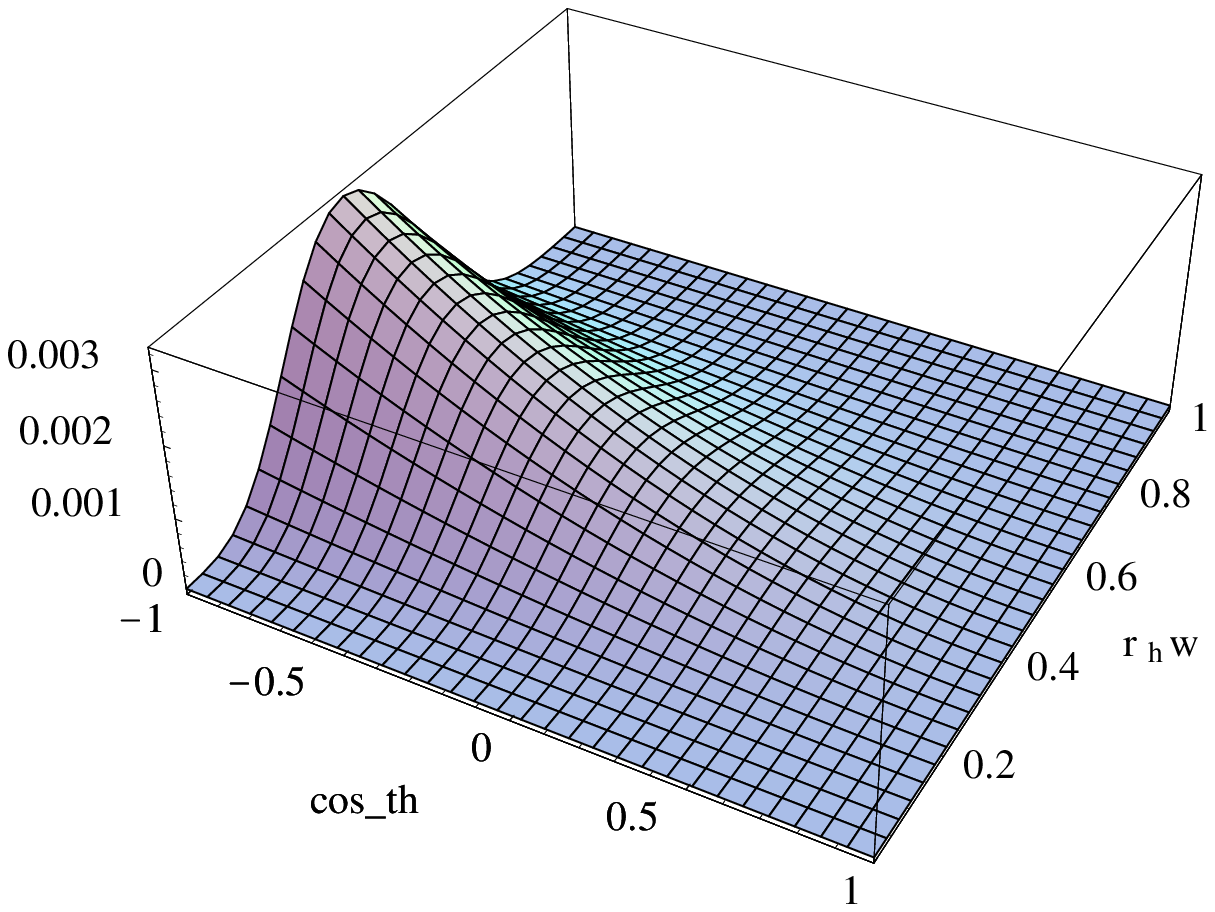,width=\regularwidthofpicture}
	\caption{
	  Vector ($s=1$) power spectrum $r_hdE/dt\,d\omega\,d\cos\vartheta$
	  vs $r_h\omega$ and $\cos\vartheta$ for $\astar=1.5$ \label{fig:ang1} } 
    \end{center}
  \end{figure}

We can see that the spectrum is substantially different from the g.o.\ limit.
(The low frequency behavior of vector emission is different
from the g.o.\ limit in its powers of $\omtil$;
see Fig.~\ref{fig:logs1} in the log-log plot.)
When the black hole is highly rotating
(\textit{i.e.} when $\astar$ is large),
the power spectrum is substantially reduced from the g.o.\ limit
especially for scalars and spinors ($s=0$ and $\frac{1}{2}$).
This can be considered as super\-radiant effect
enhancing the higher spin emission ($s=1$)
compared to the lower ones ($s=0$ and $\frac{1}{2}$).
(This can also be seen by comparing the height of the peaks
in Figs.~\ref{fig:ang12} and \ref{fig:ang1}.)
We expect that this super\-radiance will be more
significant when $n$ is large
since $(\astar)_\mathrm{max}={n+2\over 2}$ is much larger
for, say, $n=7$ than for $n=1$
making the higher powers of $\Qtil$ ($\simeq -m\astar$)
more significant in Eq.~\eqref{eq:Gamma_explicit}
(or in its counterpart for $n\geq 2$).

\begin{figure}[tbhp]
  \begin{center}
    \epsfig{file=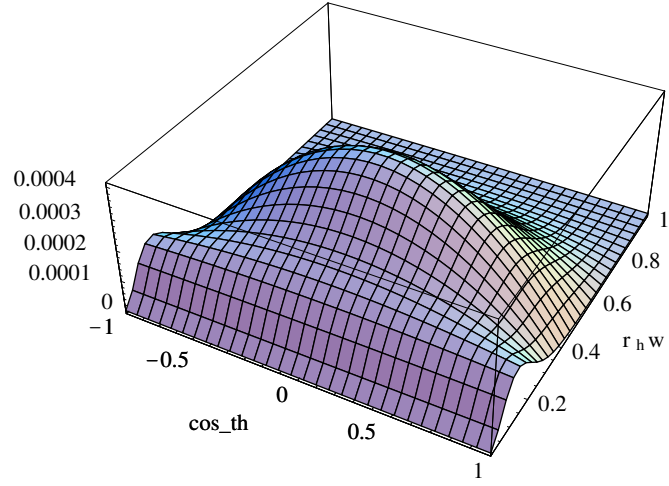,width=\regularwidthofpicture}
    \caption{
	  Scalar ($s=0$) power
      spectrum $r_hdE/dt\,d\omega\,d\cos\vartheta$ vs $r_h\omega$ and
      $\cos\vartheta$ for $\astar=1.5$ \label{fig:ang0} } 
    \end{center}
  \end{figure}

In Figs.~\ref{fig:ang0}--\ref{fig:ang1}, we present the angular
dependent power spectra for spin $s=0$, $\frac{1}{2}$ and 1
fields when $\astar=(\astar)_\mathrm{max}=1.5$. 
Recall that our approximation is valid for $\omtil<\min(1,\astar^{-1})$.
We observe that there is a large angular dependence.
Note that $\cos\vartheta=1$ ($-1$)
is the direction (anti)parallel to the angular momentum of the black
hole which is perpendicular to the beam axis
and that $\cos\vartheta=0$ contains the direction
of beam axis (for some value of $\varphi$).
(Even after averaging over the rotation around the beam axis,
there still remains strong angular dependences for spinor and vector
fields.)
We note that since only one helicity component is used
in Figs.~\ref{fig:ang12} and \ref{fig:ang1},
the large up-down asymmetry does not imply a parity violation
when we sum up the opposite components.
(Neutrinos are exceptional since
they do not have opposite helicity states to be paired with them.)
The angular
dependence shown in Figs.~\ref{fig:ang0}--\ref{fig:ang1} vanishes
when we take the limit $\astar\rightarrow 0$.

\section{Summary and discussion}
We have shown that black holes tend to be produced with large
angular momentum
and that the production cross section of the black hole 
increases when one takes this into account.
We have also estimated the possibility of black ring formation
and found that it may be produced when there are many extra dimensions,
though its fate is currently unknown.
It would be interesting to estimate the black ring 
production cross section too. (One way might be to follow our argument
for the black hole production.)

We have presented the spin $s=0$, $\frac{1}{2}$ and 1 brane field equations
for the  $(4+n)$-dimensional Kerr black hole and have shown that they are separable.
We have analytically solved them for the five dimensional case ($n=1$) and obtained the
greybody factor in the low frequency limit.
Note that
our results include the case of
\Sch black hole in the limit $a\rightarrow 0$.
This is the first time that
the greybody factors of brane spinor and vector fields
are obtained for the higher dimensional black hole
regardless of whether it is rotating or not.
We have found that the resulting power spectra are substantially
modified from the geometrical optics limit
and that there is strong angular dependence.
We expect that these features remain qualitatively the same
for any number of extra dimensions $n$.

It is important to obtain the greybody factors
for higher dimensional Kerr black hole
numerically
without relying on the low frequency expansion to determine
the total radiation of the black hole from its birth till the end
for general $n$.
This work is in progress.

\section*{Acknowledgments}
We are grateful to the organizers and participants
of the 10th Marcel Grossmann Meeting in Rio de Janeiro,
20--26 July 2003
for hosting and exchanging stimulating discussions.
We thank Panagiota Kanti for the discussion summarized in Appendix
and Manuel Drees for reading the manuscript.
The work of K.O.\ is partly supported by the SFB375
of the Deutsche Forschungs\-gemein\-schaft.

\appendix

\section{Brane field equations}
\renewcommand{\theequation}{A.\arabic{equation}}
In Ref.~\mycite{Kanti:2002ge}
which appeared after our work,
the brane field equation 
for the higher dimensional \Sch black hole
(\textit{i.e.} the limit $a\rightarrow 0$ in our language),
\begin{align}
  &\Delta^s{d\over dr}\left[\Delta^{1-s}{dR_s\over dr}\right]
  +\Biggl\{
    {\Sigma^2\omega^2-is\omega\Sigma\partial_r\Delta\over\Delta}
    +2is\omega\partial_r\Sigma-\Lambda \Biggr. \nonumber\\
  &\qquad\left.
    +\Delta\left(s-\frac{1}{2}\right)
	  \left[\partial_r\left(\partial_r\Sigma\over\Sigma\right)
	  +\left(s-\frac{1}{2}\right)\left(\partial_r\Sigma\over\Sigma\right)^2
	  +(1-s){\partial_r\Sigma\over\Sigma}{\partial_r\Delta\over\Delta}
	  \right]
    \right\}R_s=0, \label{KM_eq}
  \end{align}
with $\Lambda=j(j+1)-s(s-1)$ being the angular eigenvalue and $\Sigma=r^2$,
is derived from the Cvetic-Larsen equation
which essentially relies on the fact that $\Delta_{,rr}-2$ vanishes
in four dimensions ($n=0$), though not in higher dimensions ($n\geq 1$).

From Eq.~\eqref{brane_field_equation_R} we can redefine the radial
function as $R=\Delta^s R_{KM}$ to obtain
\begin{align}
  &\Delta^s{d\over dr}\left[\Delta^{1-s}{dR_{KM}\over dr}\right]\nonumber\\
  &\qquad+\left[{K^2-isK\Delta_{,r}\over\Delta}+4isr\omega-2s-A+2ma\omega-(a\omega)^2\right] R_{KM}
  =0. \label{from_ours}
  \end{align}
In the limit $a\rightarrow 0$,
$K$ and the angular eigenvalue $A$ behave as
\begin{align}
  K &\rightarrow r^2\omega=\Sigma\omega, &
  A &\rightarrow l(l+1)-s(s+1)=\Lambda-2s,
  \end{align}
and therefore Eq.~\eqref{KM_eq} becomes identical to Eq.~\eqref{from_ours}
when one deletes the second line of Eq.~\eqref{KM_eq}:
\begin{align}
  \Delta\left(s-\frac{1}{2}\right)
	  \left[\partial_r\left(\partial_r\Sigma\over\Sigma\right)
	  +\left(s-\frac{1}{2}\right)\left(\partial_r\Sigma\over\Sigma\right)^2
	  +(1-s){\partial_r\Sigma\over\Sigma}{\partial_r\Delta\over\Delta}
	  \right]
    &\longrightarrow 0. \label{correction_eq}
  \end{align}
(The l.h.s.\ of Eq.~\eqref{correction_eq} becomes
zero for $s=\frac{1}{2}$ and 1, though not for $s=0$,
and hence the result of Ref.~\mycite{Kanti:2002ge}
studying the former cases
remains unchanged after this correction.)
This check is first presented in Ref.~\mycite{Harris:2003eg}
by utilizing our Newman-Penrose tetrads~\eqref{eq:null_tetrad}
and by rederiving the the master equation~\eqref{brane_field_equation_R}
both in the limit $a\rightarrow 0$.

\bibliography{paper}
\bibliographystyle{utphys}

\end{document}